\documentclass[10pt,letterpaper]{article}
\usepackage{authblk}

\usepackage[top=0.85in,bottom=1in,left=1in,right=1in,footskip=0.75in]{geometry}

\usepackage{amsmath,amssymb}
\usepackage{graphicx}
\usepackage{bm}
\usepackage{cite}
\usepackage{comment}

\def\eps{{\varepsilon}}

\title{An Isochron-Free Framework for Phase Reduction and Coupling Inference}

\author[1,2]{Akari Matsuki \thanks{amatsuki@ictp.it}}
\author[3,4]{Ryota Kobayashi}
\author[3]{Hiroshi Kori \thanks{kori@k.u-tokyo.ac.jp}}

\affil[1]{Quantitative Life Sciences, The Abdus Salam International Centre for Theoretical Physics, Trieste, Italy}

\affil[2]{Faculty of Advanced Life Science, Hokkaido University,  Sapporo, Japan}

\affil[3]{Graduate School of Frontier Sciences, The University of Tokyo, Chiba, Japan}

\affil[4]{Mathematics and Informatics Center, The University of Tokyo, Tokyo, Japan }

\begin{document}

\maketitle


\begin{abstract}
Phase description provides a compact and powerful framework for analyzing synchronization dynamics in weakly coupled limit-cycle oscillators. While its classical formulation relies on the asymptotic phase defined by isochrons, reconstructing isochrons from observed trajectories is often challenging for complex models and real-world systems. Here we develop an isochron-free framework based on a readily computable generalized phase, such as the polar angle computed from observed trajectories. We theoretically show that, under near-uniform rotation of the generalized phase and sufficiently stable amplitude dynamics, a one-period stroboscopic description yields a closed circle map. The interaction term of the resulting circle map coincides, to leading order, with the phase coupling function obtained from the conventional phase reduction. Based on this circle map, we propose a method to infer coupling from oscillatory time series. The method is validated using synthetic data from van der Pol oscillators. Our framework broadens the applicability of phase reduction and provides a theoretically grounded method for coupling inference from oscillatory data.
\end{abstract}

\section{Introduction}
Coupled limit-cycle oscillators provide a unified, simple description of rhythmic phenomena across physics, chemistry, biology, and engineering, where collective behaviors such as synchronization and entrainment emerge from interactions among self-sustained oscillations \cite{Kuramoto1984, Winfree2001, Nakao2014PRX}.
Phase reduction is a central tool in this area: it reduces high-dimensional oscillator dynamics to low-dimensional phase descriptions, enabling tractable analysis of phase locking, stability, and collective behavior under weak perturbations \cite{Guckenheimer1975, Winfree2001, SchwemmerLewis2012, kuramoto1984chemical, hoppensteadt1997}.

Classical phase reduction relies on the \emph{asymptotic phase} defined as a specific function of the state vector, which yields a closed phase-only dynamics near a stable limit cycle \cite{Guckenheimer1975, Winfree2001}.
In practice, however, constructing the asymptotic phase often challenging when only limited data are available and generally requires substantial computation \cite{namura2022estimating, yawata2024phase, yamamoto2025gaussian,koyama2026}.  
This motivates the use of practically accessible phase-like coordinates.
When a multidimensional state vector is available, one may use a smooth state-dependent function, such as a polar angle. We refer to this type of phase \emph{generalized phase}. 
It is also common to construct phase-like variables with the aid of delay coordinates \cite{packard1980geometry, takens1981detecting, rosenblum1996phase, pikovsky2003synchronization} or the Hilbert transform \cite{delprat1992, cohen1999ambiguity, chavez2006towards}, especially when only one-dimensional signal can be observed.

However, a fundamental obstacle to using generalized phases is that, in continuous time, the phase dynamics are generally contaminated by amplitude deviations. 
In this study, we show that this issue can be resolved by shifting from continuous-time descriptions to a \emph{stroboscopic} (one-period) representation. 
Using Floquet theory~\cite{SchwemmerLewis2012,ErmentroutParkWilson2020arXiv,ParkErmentrout2016JCN}, 
we show that amplitude deviations become effectively slaved to the phases after a rapid transient, provided the amplitude stability is sufficiently strong and the chosen generalized phase rotates approximately uniformly along the unperturbed cycle. 
As a consequence, we obtain a \emph{circle map}, which describes the period-to-period dynamics of the phase. 
Remarkably, the coupling is characterized by exactly the same coupling function as that obtained from the phase reduction. 
This provides an \emph{isochron-free} framework for phase reduction using readily comptable phase coordinates. 

A second motivation for this study is the inference of coupling functions from oscillatory time-series data. 
While the asymptotic phase is usually difficult to construct, the generalized phase is often feasible to define. 
We therefore propose a coupling-inference method based on generalized phases and the circle map. 
A key property of the circle map is that the coupling function appearing in the map coincides with that obtained from the conventional phase reduction.  
This implies that the inferred coupling function is independent of the particular choice of phase.
We demonstrate the effectiveness of the proposed method using simulated data, and furhter provide numerical evidence that the method remains effective even when only a one-dimensional observable is available.
This work extends a previous study \cite{matsuki2025network}, which established the inference method for asymptotic phases, to a broader class of generalized phases.

\paragraph{Organization of the paper.}
Section~\ref{theory} introduces asymptotic phase and generalized phases, and provides the theory for describing their dynamics. 
Section~\ref{derivation} presents the derivation of a closed phase-difference circle map under the stated assumptions for two coupled oscillators.
Section~\ref{extensions} briefly discusses the extension of the theory to networks of oscillators and higher-order interactions. 
Section~\ref{sec:couling-inf} applies the resulting circle map to coupling inference from synthetic data. In Sec.~\ref{conclusion}, we conclude the study and discuss future directions.

\section{Theory}
\label{theory}
\begin{figure}
    \centering
    \includegraphics[width=0.8\linewidth]{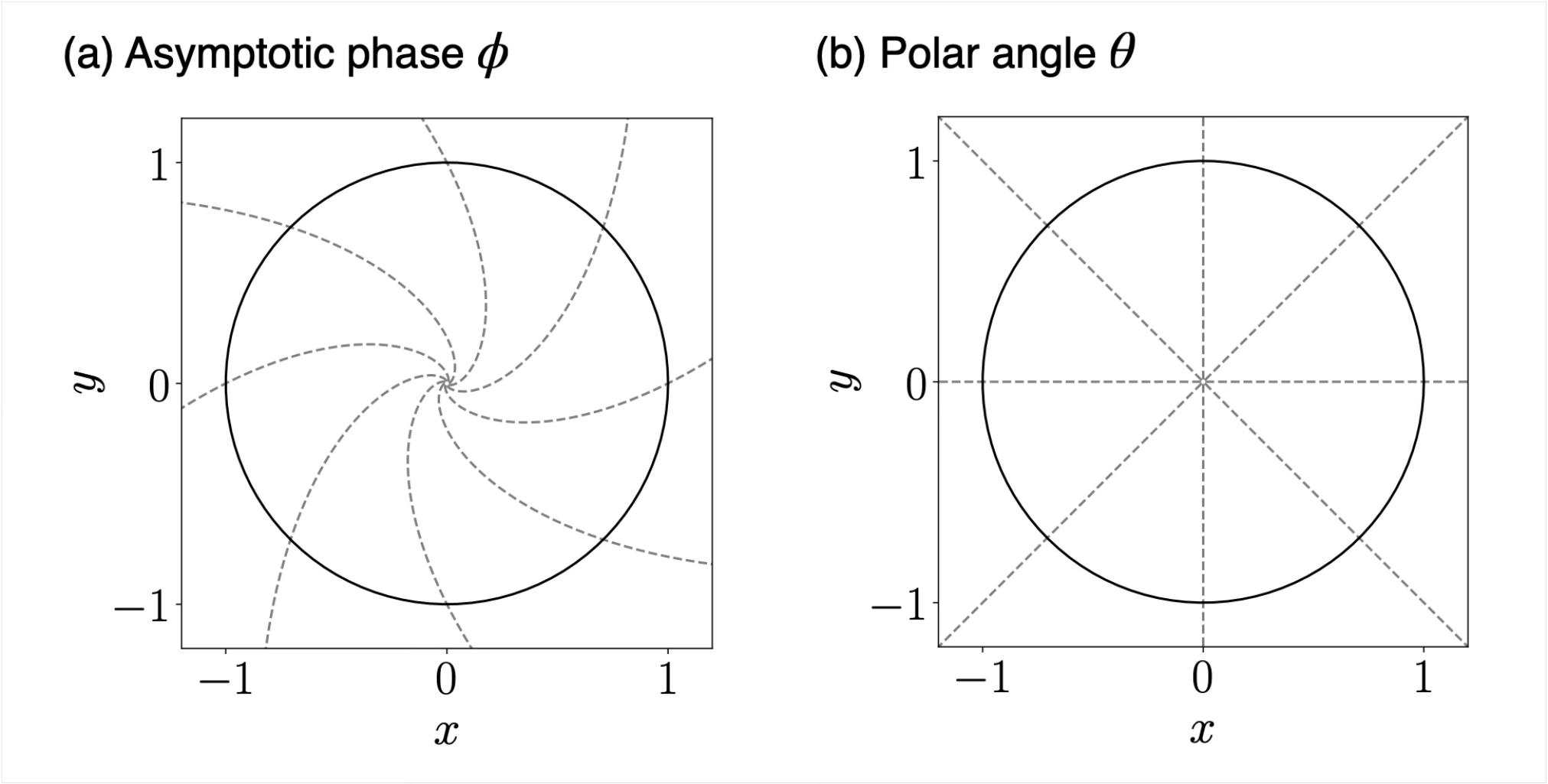}
    \caption{(a) Asymptotic phase and (b) polar angle as an example of generalized phase. The solid and dashed lines represent the limit cycle and the isophase lines, respectively, generated using the Stuart-Landau oscillator with $\alpha=0.5$ [Eq.~\eqref{sl}]. Note that the isophase lines of the asymptotic phase, i.e., isochrons, are model/parameter-dependent. }
    \label{fig1-phases}
\end{figure}

This section provides a roadmap to the derivation in Sec.~3 and fixes notation.
We first recall the asymptotic phase (See Fig.\ref{fig1-phases}a as an example), which is defined in a system-dependent manner, and review the standard phase reduction. We demonstrate that the asymptotic-phase equation closes at leading order and yields a phase-difference circle map under weak coupling. 
We then introduce a \emph{generalized phase}, defined as a smooth function of the system state (e.g., a polar angle from full-state observation; See Fig.\ref{fig1-phases}b), and explain why its continuous-time dynamics is generally \emph{not} closed due to $O(\varepsilon)$ amplitude contamination.
The central idea of this paper is that the obstruction can be removed at the \emph{stroboscopic} (one-period) level once the amplitude dynamics is properly accounted for. 

Section \ref{dyn-coupled-lc} provides the dynamical description of weakly coupled limit-cycle oscillators. In Sec. \ref{asympt-phase-reduction}, we present the classical phase reduction theory based on the asymptotic phase. In Sec. \ref{gen-phase}, we define the generalized phase and demonstrate the challenges associated with its use. 
Section \ref{main-result} provides the main result, namely, the circle map for the generalized phase, and its implications. 
Finally, in Sec. \ref{main-result}, we numerically verify that the circle map describes the dynamics of the generalized phase more accurately than the classical continuous-time model.   

\subsection{Dynamical description of coupled limit-cycle oscillators}
\label{dyn-coupled-lc}
We consider a limit-cycle oscillator described by
\begin{align}
\label{oscillator}
    \frac{d\bm X}{dt} = \bm F(\bm X),
\end{align}
where \(\bm X(t)\in\mathbb{R}^d\) denotes the state vector and
\(\bm F:\mathbb{R}^d\to\mathbb{R}^d\) is a smooth vector field.
For convenience, we rescale time so that the period of the stable limit cycle is $T=2\pi$.
Throughout, we restrict attention to the basin of attraction of the limit cycle, where phase coordinates are well defined.
We refer to~\eqref{oscillator} as the \emph{unperturbed system}.

We then consider the dynamics of the oscillator weakly coupled to another, weakly nonidentical oscillator:
\begin{subequations}
 \label{oscillators}
\begin{align}
    \frac{d\bm X}{dt} &= \bm F(\bm X) + \varepsilon\, \bm P(\bm X,\bm X'), \label{osci1}\\
    \frac{d\bm X'}{dt} &= \bm F(\bm X') + \varepsilon \Delta \boldsymbol{F}(\bm X') + \varepsilon\, \bm P'(\bm X',\bm X). \label{osci2}
\end{align}
\end{subequations} 
Here, $\bm X(t)$ denotes the state of the oscillator of interest, while $\bm X'(t)$ denotes the state of the other oscillator.
The parameter $\varepsilon$, with $0<\varepsilon\ll1$, is a small parameter characterizing both the strengths of coupling and heterogeneity.
The terms $\eps \bm P(\cdot,\cdot)$ and $\eps \bm P'(\cdot,\cdot)$ describe the couplings, and $\varepsilon \Delta \bm F(\cdot)$ represents the heterogeneity between the oscillators.
The functions $\bm F(\cdot),\Delta \bm F(\cdot), \bm P(\cdot,\cdot)$ and $\bm P'(\cdot,\cdot)$ are assumed to be $O(1)$.
Hence, the intrinsic dynamics occur on an $O(1)$ time scale, while the coupling and heterogeneity enter as $O(\varepsilon)$ perturbations.

The primary goal of phase reduction is to obtain a closed, low-dimensional description of~\eqref{oscillators} in terms of phase variables only, ideally as a two-dimensional system on the torus (or, equivalently, a one-dimensional description for the phase difference).

\subsection{Asymptotic phase and the standard phase reduction}
\label{asympt-phase-reduction}
Following the classical theory of isochrons and asymptotic phase
\cite{winfree2001geometry, Nakao2016}, we introduce a phase
function $\Phi$ in the basin of attraction of the stable limit cycle.
For each point $\bm X$ in the basin, $\Phi(\bm X)$ assigns the phase
of the point on the limit cycle to which the trajectory starting from
$\bm X$ asymptotically converges. 

Letting $\phi=\Phi(\bm X)$ and using $T=2\pi$,
the phase function satisfies
\begin{align}
 \frac{\partial \Phi}{\partial \bm X}\cdot \bm F(\bm X)=1,
\label{asymptotic_phase}
\end{align}
so that along trajectories of the unperturbed system~\eqref{oscillator},
\begin{align}
  \frac{d\phi}{dt}
  &= \frac{d}{dt}\Phi(\bm X(t))
   = \frac{\partial \Phi}{\partial \bm X}\cdot \frac{d\bm X}{dt}
   = \frac{\partial \Phi}{\partial \bm X}\cdot \bm F(\bm X(t))
   = 1 .
\end{align}
Thus, the asymptotic phase advances at constant speed everywhere in the basin, not only on the limit cycle.
We fix the phase origin so that $\phi=0$ corresponds to a reference point $\bm X_0(0)$ on the limit cycle.
With this convention, the limit-cycle trajectory $\bm X_0(t)$ satisfies $\phi(t)=\Phi(\bm X_0(t))=t$. 

For the perturbed system~\eqref{oscillators}, the phase of oscillator 1, $\phi(t)=\Phi(\bm X(t))$, evolves as
\begin{align}
  \frac{d\phi}{dt}
  &= \frac{\partial \Phi}{\partial \bm X}\cdot \frac{d\bm X}{dt}
   = \frac{\partial \Phi}{\partial \bm X}\cdot \bm F(\bm X(t))
     +\varepsilon\,\frac{\partial \Phi}{\partial \bm X}\cdot \bm P(\bm X,\bm X') \notag\\
  &= 1+\varepsilon\, \bm Z(\phi)\cdot \bm p(\phi,\phi')+O(\varepsilon^2),
  \label{dphi_dt}
\end{align}
where $\phi'=\Phi(\bm X')$ and
\begin{align}
 \bm Z(\phi) &= \left.\frac{\partial \Phi}{\partial \bm X}\right|_{\bm X=\bm X_0(\phi)}, \\
 \bm p(\phi,\phi') &= \bm P(\bm X_0(\phi),\bm X_0(\phi')).
\end{align}
At leading order, the dynamics closes in terms of phases.

A stroboscopic representation is obtained by integrating \eqref{dphi_dt} over one period.
Introducing $\psi=\phi-t$ and $\psi'=\phi'-t$, Eq.~\eqref{dphi_dt} reduces to
\begin{align}
 \frac{d\psi}{dt} =  \varepsilon\, \bm Z(\psi+t)\cdot \bm p(\psi+t,\psi'+t)+O(\varepsilon^2).
 \label{dpsi_dt}
\end{align}
To integrate this equation over $(t_0, t_0+2\pi)$,
we substitute $\psi(t) = \psi(t_0) + O(\eps)$, which is valid for $t-t_0 = O(1)$, into Eq.~\eqref{dpsi_dt} to obtain
\begin{align}
 \frac{d\psi}{dt} =  \varepsilon\, \bm Z(\psi(t_0)+t)\cdot \bm p(\psi(t_0)+t,\psi'(t_0)+t)+O(\varepsilon^2).
 \label{dpsi_dt2}
\end{align}
Now, the right-hand side of the equation is $2\pi$-periodic in $t$ at the leading order. By integration, we obtain
\begin{align}
\frac{\phi(t_0+2\pi)-\phi(t_0)}{2\pi}
= 1+\varepsilon\,\Gamma\bigl(\phi(t_0)-\phi'(t_0)\bigr)
+O(\varepsilon^2).
\label{map_ph_tmp}
\end{align}
A more symmetric expression may be obtained by replacing 
$t_0$ by $t-\pi$ and using $\Gamma\bigl(\phi(t-\pi)-\phi'(t-\pi)\bigr) = \Gamma\bigl(\phi(t)-\phi'(t)\bigr) + O(\eps)$ in Eq.~\eqref{map_ph_tmp}, resulting in
\begin{align}
\frac{\phi(t+\pi)-\phi(t - \pi)}{2\pi}
= 1+\varepsilon\,\Gamma\bigl(\phi(t)-\phi'(t)\bigr)
+O(\varepsilon^2),
\label{map_ph}
\end{align}
with the coupling function
\begin{align}
 \Gamma(\phi-\phi')
 =\int_0^{2\pi} ds\;
 \bm Z(s+\phi)\cdot \bm p(s+\phi,\,s+\phi'). \label{Gamma}
\end{align}
Importantly, the interaction term depends only on the phase difference.

\subsection{Generalized phase and the non-closure issue}
\label{gen-phase}
In applications, one often works with phase coordinates that are easier to construct than the asymptotic phase.
In this paper, we use the term \emph{generalized phase} to mean a phase-like coordinate defined as a smooth function of the \emph{state}.
Let
\begin{align}
\theta=\Theta(\bm X)
\end{align}
be such a generalized phase, assumed to be topologically conjugate to the asymptotic phase on and near the limit cycle.
Parameterizing the limit cycle by $\theta$, we write
\begin{align}
    \tilde{\bm X}_0(\theta)=\bm X_0(t),
\end{align}
where $\theta=\Theta(\bm X_0(t))$.
An example is a polar angle (e.g., the polar angle in a planar projection), but many other state-dependent definitions are possible, as exemplified later.

Differentiating $\theta(t)=\Theta(\bm X(t))$ along the trajectories of the purturbed system \eqref{oscillators} yields
\begin{align}
  \frac{d\theta}{dt}
  &= \frac{\partial \Theta}{\partial \bm X}\cdot \frac{d\bm X}{dt} \\
  &= \frac{\partial \Theta}{\partial \bm X}\cdot \bm F(\bm X(t))
     +\varepsilon\,\frac{\partial \Theta}{\partial \bm X}\cdot \bm P(\bm X,\bm X') \notag\\
  &= \left[\tilde{\bm Z}(\theta)+O(\varepsilon)\right]\cdot
     \left[\tilde{\bm f}(\theta)+O(\varepsilon)\right]
    +\varepsilon\, \tilde{\Gamma}(\theta, \theta')
    +O(\varepsilon^2), \label{dot_theta_Z}
\end{align}
where
\begin{align}
\tilde{\bm Z}(\theta)
&=\left.\frac{\partial \Theta}{\partial \bm X}\right|_{\bm X=\tilde{\bm X}_0(\theta)},\\
\tilde{\bm f}(\theta)
&=\bm F\bigl(\tilde{\bm X}_0(\theta)\bigr),\\
\tilde{\Gamma}(\theta, \theta') &= \tilde{\bm Z}(\theta)\cdot \tilde{\bm p}(\theta,\theta'),\\
\tilde{\bm p}(\theta,\theta')
&=\bm P\bigl(\tilde{\bm X}_0(\theta),\,\tilde{\bm X}_0(\theta')\bigr).
\end{align}
We see that $O(\eps)$ terms appear in Eq.~\eqref{dot_theta_Z}. This is because, unlike the asymptotic phase, a generalized phase $\Theta(\bm X)$ generally \emph{does not} satisfy the identity \eqref{asymptotic_phase}.
The $O(\varepsilon)$ terms generally depend on amplitude deviations from the cycle, and thus, the generalized-phase equation is not closed at leading order in continuous time.
This is the main obstacle in using non-isochronal, state-defined phase coordinates as replacements for the asymptotic phase in standard phase reduction.

Moreover, the continuous-time description retains a limitation that is already present for the asymptotic phase: the coupling term is generally not a function of a single phase difference, but rather a function of the two phases $\theta$ and $\theta'$ separately.

\subsection{Main result and its implication}
\label{main-result}
We show that the non-closure in continuous time can be resolved at the stroboscopic level under two practically feasible assumptions:
\begin{enumerate}
\item[(i)] The amplitude relaxation is sufficiently strong compared with the coupling, so that amplitude deviations decay on an $O(1)$ timescale while the coupling acts on an $O(\varepsilon^{-1})$ timescale.
\item[(ii)] The generalized phase rotates nearly uniformly on the unperturbed limit cycle, i.e., along $\bm X_0(t)$ it satisfies $\dot\theta = 1 + O(\varepsilon)$.
\end{enumerate}
Under these assumptions, amplitude deviations become effectively slaved to the phases after a short transient, and the one-period update of the generalized phase admits a closed circle-map representation:
\begin{align}
 \frac{\theta(t+\pi)-\theta(t - \pi)}{2\pi}
 = 1 + \varepsilon\, \Gamma\!\left(\theta(t)-\theta'(t)\right)
 + O(\varepsilon^2).
 \label{map_th}
\end{align}
This equation is exactly the same as that for the asymptotic phase up to $O(\eps)$, Eq.~\eqref{map_ph}; importantly, the coupling term depends only on the phase difference and is described by the same function $\Gamma$.

This result is valuable in practice.
While constructing the asymptotic phase (isochrons) can be difficult, it is often feasible to define a state-based generalized phase that evolves approximately uniformly on the limit cycle.
For example, a polar angle from two-state observations can have this property for some oscillatory systems.
Once such a phase is available, the resulting circle map retains the phase-difference structure, yielding a concise model.
Moreover, in Sec.~\ref{sec:couling-inf}, we will numerically demonstrate that the method is effective even when only a single observable is available by employing a delay coordinate.

\subsection{Numerical verification}
\label{numerical-verification}
In the context of oscillatory data analysis, it is common to assume the continuous-time phase model such as
\begin{align}
    \frac{d\theta}{dt} = 1 +  \eps \Gamma (\theta(t) - \theta'(t)).
    \label{contin-model}
\end{align}
We examine how accurately these two models \eqref{map_th} and \eqref{contin-model} describe the phase dynamics of the coupled Stuart-Landau oscillators. 
\begin{subequations}\label{sl}
\begin{align}
    \frac{d x_i}{dt} &= x_i - (\omega_i + \alpha) y_i - (x_i^2 + y_i^2) \left( x_i - \alpha y_i\right) + \eps c_i \left(x_j - x_i \right), \\
    \frac{d y_i}{dt} &= (\omega_i + \alpha) x_i + y_i - (x_i^2 + y_i^2) (y_i + \alpha x_i),
\end{align}
\end{subequations}
where $i, j \in \{1,2 \} \ (i \neq j)$ are the indices of the oscillators. 
As an example of a generalized phase, consider the polar angle (Fig.~\ref{fig1-phases}b)
\begin{align}
    \theta_i = \arctan\left[ \frac{y_i}{x_i} \right].
\end{align}
Note that, for the Stuart-Landau oscillator, the isochron can analytically be derived and the asymptotic phase is described as (Fig.~\ref{fig1-phases}a)
\begin{align}
    \phi_i = \theta_i - \alpha \ln \sqrt{x_i^2 + y_i^2}, \label{sl-phi-theta}
\end{align}
which is different from the asymptotic phase when $\alpha \neq 0$ .

\begin{figure}
    \centering
    \includegraphics[width=\linewidth]{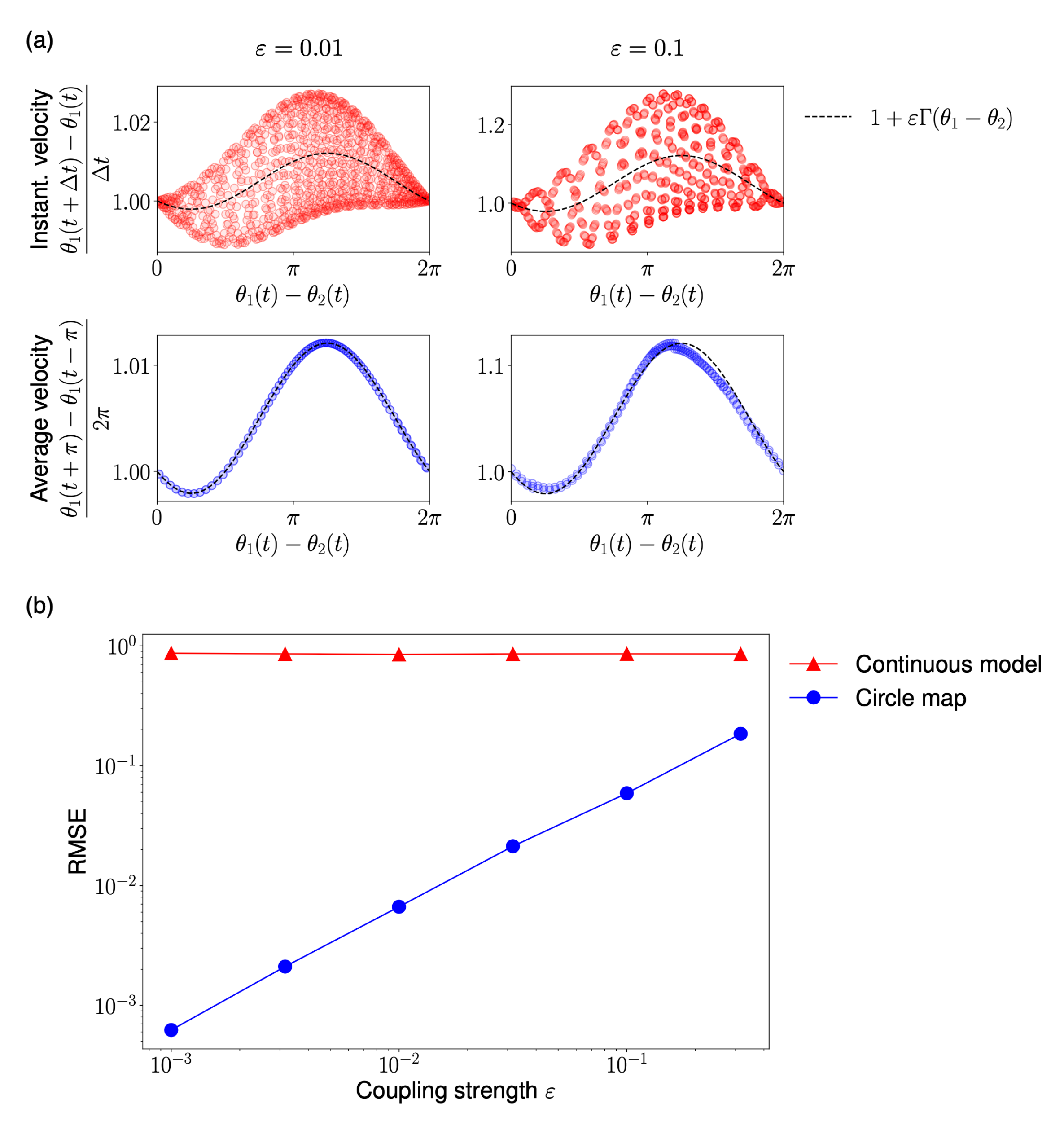}
    \caption{The circle map describes the dynamics of the generalized phase more accurately than the continuous-time model. 
    (a) The instantaneous phase velocity as a function of phase difference $\theta_1 - \theta_2$ (red dots) and the average phase velocity (blue dots), with the analytically obtained $1 + \eps \Gamma_1(\theta_1 - \theta_2)$ (black dashed lines).
    (b) The deviation between the model and the data as a function of $\eps$. 
    Parameters: $\omega_1=1.0, \omega_2 = \omega_1 + 2\eps, \ \alpha= 1.0, \ c_1=1, c_2=0$.}
    \label{fig2}
\end{figure}

According to the circle map \eqref{map_th}, $1 + \eps \Gamma(\theta_1(t) - \theta_2(t))$ corresponds to the average phase velocity over one period
\begin{align}
    v_{\rm avr}(t) := \frac{\theta_1(t+\pi) - \theta_1( t-\pi)}{2\pi}.  \label{v_T}
\end{align}
In contrast, the continuous model \eqref{contin-model} identifies $1 + \eps \Gamma(\theta_1(t) - \theta_2(t))$ with the instantaneous phase velocity 
\begin{align}
    v_{\rm inst}(t) := \frac{d \theta_1(t)}{dt}. \label{v_dt}
\end{align}
We examine the accuracy of these two descriptions with numerical data. 
Figure \ref{fig2}a shows $1 + \eps \Gamma(\theta_1 - \theta_2)$ (black dashed line), where $\Gamma(\psi) = 1 + \frac{\eps}{2} \left( - \sin \psi - \cos \psi + 1 \right)$ is the coupling function of the asymptotic phase equation, and the data of the instantaneous velocity (red circles) and that of the average velocity over one period (blue circles). 
Note that the instantaneous velocity is approximated by
\begin{align}
    v_{\rm inst}(t) \approx \frac{\theta_1(t+\Delta t) - \theta_1(t)}{\Delta t},    
\end{align}
where $\Delta t= 0.01$ is a small time step. 
The instantaneous velocity substantially deviates from the black dashed curve, while the average velocity agrees closely with it. 
This means that the circle map achieves a better fit to the data than the contiuous model. 
Importantly, it also means that the coupling function of the circle map for the polar angle coincides with the coupling function $\Gamma$ of the asymptotic-phase equation.
Figure \ref{fig2}b quantifies the deviation between the model and the data as a function of the coupling strength $\eps$. 
As we want to focus on the asynchronous regime, we scale the difference in the natural frequency with $\eps$, setting $\omega_2 = \omega_1 + 2\eps$ while fixing $\omega_1=1$. 
We define the deviation by a rescaled root-mean-square-error
\begin{align}
    {\rm RMSE} = \sqrt{\langle (\varepsilon^{-1}(1 + \eps \Gamma(\theta_1 - \theta_2) - v(t) ))^2 \rangle} , 
\end{align}
where $\langle \ \rangle$ is the average over data points. The phase velocity is $v(t) = v_{\rm inst}(t)$ \eqref{v_dt} for the continuous model and  $v(t) = v_{\rm avr}(t)$ \eqref{v_T} for the circle map.
The RMSE associated with the continuous-time model remains of $O(1)$. 
This indicates that the continuous model \eqref{contin-model} contains an error of $O(\eps)$, as emphasized in \cite{matsuki2025network}.
On the other hand, the RMSE of the circle map is of $O(\eps)$, and 
this scaling agrees with the theoretical error $O(\eps^2)$ of  \eqref{map_th}. 


\section{Derivation}
\label{derivation}
This section presents the derivation of the circle map \eqref{map_th}. Readers primarily interested in the applications may safely skip the technical details. 
The derivation also clarifies why the two assumptions introduced in Sec.~\ref{main-result} are required and explains why considering a stroboscopic map allows the description involving only phase variables.  

To keep the notation simple, we assume that the oscillator state is two-dimensional, i.e., $\bm X(t)\in\mathbb{R}^2$.
All arguments below extend to higher dimensions as long as the limit cycle is hyperbolic with a single neutral direction and the remaining Floquet exponents have negative real parts.

\subsection{Amplitude dynamics}
We first show that, to leading order in the coupling strength $\varepsilon$, the amplitude deviation from the limit cycle can be expressed as a function of the (asymptotic) phases.
This is an adiabatic elimination in the sense that the transverse (amplitude) mode relaxes rapidly compared with the slow phase drift induced by weak coupling.
Our derivation uses Floquet theory for the linearization around the limit cycle together with a perturbation expansion in $\varepsilon$.

Consider the coupled system~\eqref{oscillators}.
Let the asymptotic phases be
\begin{align}
\phi(t)=\Phi(\bm X(t)), \qquad \phi'(t)=\Phi(\bm X'(t)),
\end{align}
and denote the initial phases by
\begin{align}
\phi_0=\phi(0), \qquad \phi_0'=\phi'(0).
\end{align}
(Here and below, the prime indicates quantities associated with oscillator~2.)

We introduce the deviations from the unperturbed limit-cycle orbit $\bm X_0(t)$ as
\begin{subequations}\label{z}
\begin{align}
\bm z(t) &= \bm X(t)-\bm X_0(t+\phi_0),\\
\bm z'(t) &= \bm X'(t)-\bm X_0(t+\phi_0').
\end{align}
\end{subequations}
We assume that $|\bm z(t)|, |\bm z'(t)|=O(\varepsilon,\varepsilon|t|)$;
the potentially growing contribution in $|t|$ 
stems from the deviation along the neutral (phase) direction.
We similarly approximate
\begin{align}
\bm P(\bm X(t),\bm X'(t))
=
\bm P\!\left(\bm X_0(t+\phi_0),\,\bm X_0(t+\phi_0')\right)
+ O(\varepsilon^{n}|t|^{n+1},\varepsilon^{n}|t|^n). \quad (n=1,2,...)
\end{align}
Expanding~\eqref{osci1} to first order in $\bm z$ and $\bm z'$ yields
\begin{align}
\frac{d\bm z}{dt}
&=
L(t)\bm z
+\varepsilon\,\bm P\!\left(\bm X_0(t+\phi_0),\,\bm X_0(t+\phi_0')\right)
+ O(\varepsilon^{n+1}|t|^{n+1},\varepsilon^{n+1}|t|^n),  
\label{dot_z}
\end{align}
where $L(t)=D\bm F(\bm X_0(t+\phi_0))$ is the Jacobian matrix along the reference orbit (and similarly for the primed equation).

To separate phase and amplitude contributions, we expand $\bm z$ and $\bm z'$ in Floquet modes (see Appendix \ref{appendix-floquet} for definitions).
Let $S(t)$ be the Floquet fundamental matrix and $\bm v_0,\bm v_1$ the right Floquet eigenvectors corresponding to the neutral (phase) and stable (amplitude) directions, respectively.
We write
\begin{align}
\bm z(t)
&=
\psi(t)\,S(t+\phi_0)\bm v_0
+
a(t)\,S(t+\phi_0)\bm v_1.  \label{expansion}
\end{align}
Here, $S(t+\phi_0)\bm v_0$ is tangent to the limit cycle and equals the flow
$\bm F(\bm X_0(t+\phi_0))$.
The coefficient $\psi(t)$ represents the deviation along the phase direction (often called the \emph{detuned phase}); it is related to the asymptotic phase by $\psi(t)=\phi(t)-t$ (up to the chosen phase origin; see, e.g., \cite{Kuramoto1984}).
The scalar $a(t)$ quantifies the deviation along the stable transverse direction and will be referred to as the \emph{amplitude}.

Substituting~\eqref{expansion} into~\eqref{dot_z} and projecting onto
the phase and amplitude directions using
the left Floquet eigenvectors $\bm u_0,\bm u_1$ (see Appendix \ref{appendix-floquet} for definitions) gives
\begin{align}
\frac{d\psi}{dt}
&=
\varepsilon\,\bm u_0 \cdot \bm G(t;\phi_0,\phi_0')
+ O(\varepsilon^{n+1}|t|^{n+1},\varepsilon^{n+1}|t|^n), 
\label{dot_psi}
\\
\frac{da}{dt}
&=
-\lambda\,a
+\varepsilon\,\bm u_1 \cdot \bm G(t;\phi_0,\phi_0')
+ O(\varepsilon^{n+1}|t|^{n+1},\varepsilon^{n+1}|t|^n), 
\label{dot_a}
\end{align}
where $\lambda>0$ is the transverse stability exponent (assumed $\lambda=O(1)$), and
\begin{align}
\bm G(t;\phi_0,\phi_0')
=
S^{-1}(t+\phi_0)\,
\bm P\!\left(\bm X_0(t+\phi_0),\,\bm X_0(t+\phi_0')\right).
\label{G}
\end{align}
Note that $\bm G$ is $2\pi$-periodic in $t$ and depends on the initial phases $(\phi_0,\phi_0')$.

Solving~\eqref{dot_a} as an initial-value problem from $t=-\infty$ yields
\begin{align}
a(t;\phi_0,\phi_0')
=
\varepsilon\int_{-\infty}^{t} ds\;
e^{-\lambda (t-s)}\,
\bm u_1\cdot \bm G(s;\phi_0,\phi_0')
+O(\varepsilon^2).
\label{a}
\end{align}
The integral of the terms of $O(\varepsilon^{n+1}|t|^{n+1}, \varepsilon^{n+1}|t|^n)$
($n=1,2,\ldots$) in Eq.~\eqref{dot_a} remain
$O(\varepsilon^2)$ because
$\int_{-\infty}^{0} e^{\lambda s}|s|^n ds \propto \lambda^{-(n+1)}$ and $\lambda=O(1)$.
Importantly, $a(t)$ (and thus $\dot a(t)$ as well) are 
$O(\varepsilon)$ and $2\pi$-periodic.

\subsection{Dynamics of the generalized phase}
We next derive the 
continuous-time evolution equation for a generalized phase valid for a time interval of $O(1)$, from which
we further derive its stroboscopic evolution.
For convenience, we regard the generalized phase as a function of the phase-amplitude coordinates
$(\phi,a)$, described as
\begin{align}
\theta = \Theta(\phi,a).
\end{align}
As shown later,
to make the stroboscopic evolution closed in the generalized phase, we need to impose two assumptions on the generalized phase
restricted to the limit cycle ($a=0$):
\begin{subequations}\label{assumption}
\begin{align}
\Theta_{\phi}(\phi,0) &= 1+O(\varepsilon),
\label{assumption1}\\
\Theta_{\phi,\phi}(\phi,0) &= O(\varepsilon),
\label{assumption2}
\end{align}
\end{subequations}
where subscripts denote partial derivatives, i.e., $\Theta_{\phi} = \partial \Theta / (\partial \phi)$ and $\Theta_{\phi, \phi} = \partial^2 \Theta / (\partial \phi^2)$. 
Assumption~\eqref{assumption1} means that, on the unperturbed cycle, $\theta$ advances nearly uniformly at the same rate as the asymptotic phase (up to $O(\varepsilon)$).
Assumption~\eqref{assumption2} is a natural consequence of~\eqref{assumption1}. 
Note that this assumption does not, in general, impose smallness on $\Theta_a(\phi,0)$ or $\Theta_{\phi,a}(\phi,0)$; throughout, we regard these derivatives as $O(1)$.
See, for example, Eq. \eqref{sl-phi-theta}.

Differentiating $\theta(t)$ in time gives
\begin{align}
\frac{d \theta}{dt}
&=
\Theta_{\phi}(\phi,a)\,\frac{d\phi}{dt}
+
\Theta_{a}(\phi,a)\,\frac{da}{dt}.
\notag
\end{align}
Expanding around $a=0$ and using $\frac{d \phi}{dt}=1+\frac{d \psi}{dt}=1+O(\varepsilon)$ for $0\le t \le O(1)$, $a=O(\varepsilon)$, and $\dot a=O(\varepsilon)$, we obtain
\begin{align}
\frac{d \theta}{dt}
&=
\Theta_{\phi}(\phi,0)\, \frac{d \phi}{dt}
+
\Theta_{\phi,a}(\phi,0)\,a\,\frac{d \phi}{dt}
+
\Theta_{a}(\phi,0)\,\frac{d a}{dt}
+
O(\varepsilon^2)
\notag\\
&=
\Theta_{\phi}(\phi,0)\, \left(1+\frac{d \psi}{dt} \right)
+
\Theta_{\phi,a}(t+\phi_0,0)\,a
+
\Theta_{a}(t+\phi_0,0)\,\frac{d a}{dt}
+
O(\varepsilon^2).
\label{dot_theta}
\end{align}
Subscripts denote partial derivatives, e.g., $\Theta_{\phi,a}=\partial^2\Theta/(\partial\phi\,\partial a)$, and we used $\phi=t+\phi_0+O(\varepsilon)$ for $0\le t\le O(1)$.

By expanding the first term in Eq.~\eqref{dot_theta} in terms of $\psi-\phi_0=O(\eps)$, which is valid for $0\le t\le O(1)$, we obtain
\begin{align}
\Theta_{\phi}(\phi,0) \left(1+\frac{d \psi}{dt} \right) 
&=
\Theta_{\phi}(t+\phi_0+\psi-\phi_0,0)\, \left(1+\frac{d \psi}{dt} \right)
\notag\\
&=
\Theta_{\phi}(t+\phi_0,0)\, \left(1+\frac{d \psi}{dt} \right)
+
\Theta_{\phi,\phi}(t+\phi_0,0)\,(\psi-\phi_0)
+
O(\varepsilon^2).
\label{Theta_Phi_tmp}
\end{align}
By using the assumption \eqref{assumption}, this term is reduced to 
\begin{align}
\Theta_{\phi}(\phi,0) \left(1+\frac{d \psi}{dt} \right) =
\Theta_{\phi}(t+\phi_0,0)
+
\frac{d \psi}{dt}
+
O(\varepsilon^2).
\label{Theta_Phi}
\end{align}
Substituting~\eqref{Theta_Phi} into~\eqref{dot_theta} yields
\begin{align}
\frac{d\theta}{dt}
=
\Theta_{\phi}(t+\phi_0,0)
+
\varepsilon\,H(t;\phi_0,\phi_0')
+
O(\varepsilon^2),
\label{dot_theta2}
\end{align}
where
\begin{align}
H(t;\phi_0,\phi_0')
= \eps^{-1} \left\{ 
\frac{d \psi}{dt}
+
\Theta_{\phi,a}(t+\phi_0,0)\,a
+
\Theta_{a}(t+\phi_0,0)\,\frac{da}{dt}\right\}  .
\label{H}
\end{align}
By substituting~\eqref{dot_psi} and~\eqref{a} into~\eqref{H}, $H$ becomes an explicit $2\pi$-periodic function of $t$ for fixed $(\phi_0,\phi_0')$.
Moreover, when $H$ is integrated over one period, the second and third terms in \eqref{H} cancel each other by integration by parts. Consequently, only the first term contributes to the averaged coupling, yielding exactly the same coupling function $\Gamma$ as defined in \eqref{Gamma}. 
Integrating~\eqref{dot_theta2} over one period gives the stroboscopic map
\begin{align}
\theta(2\pi)-\theta(0)
=
2\pi\left[1+\varepsilon\,\Gamma(\phi_0-\phi_0')\right]
+O(\varepsilon^2),
\label{map-th-phi}
\end{align}
Equations~\eqref{Theta_Phi}--\eqref{map-th-phi} also clarify the role of assumption~\eqref{assumption}.
The assumption \eqref{assumption2} is crucial in the expansion \eqref{Theta_Phi_tmp}: since \(\psi-\phi_0=O(\varepsilon)\), the correction term \(\Theta_{\phi,\phi}(t+\phi_0,0)(\psi-\phi_0)\) becomes \(O(\varepsilon^2)\) and can be neglected at leading order. Consequently, the leading-order part of \eqref{dot_theta2} is \(2\pi\)-periodic in \(t\), so that integration over one period yields a stroboscopic map depending only on the phase variable.
The assumption \eqref{assumption1} implies that the first term in \eqref{H} reduces to \(d\psi/dt\) without any additional factor involving \(\Theta_\phi\). 
After period averaging, this leads to exactly the same coupling function $\Gamma$ as in the asymptotic-phase description. 
Thus, assumption~\eqref{assumption} plays a dual role: it ensures the closure of the generalized-phase dynamics in the stroboscopic map and guarantees that the resulting coupling function coincides with that of the asymptotic phase.

Finally, assumption~\eqref{assumption1} implies that, on the limit cycle,
the phase difference in $\phi$ and that in $\theta$ agree up to $O(\varepsilon)$:
\begin{align}
\phi_0-\phi_0' = \theta_0-\theta_0' + O(\varepsilon),
\label{transform}
\end{align}
where
\begin{align}
\theta_0=\Theta(\phi_0,0), \qquad \theta_0'=\Theta(\phi_0',0).
\end{align}
Substituting~\eqref{transform} into~\eqref{map-th-phi} yields
\begin{align}
\theta(2\pi)-\theta(0)
=
2\pi\left[1+\varepsilon\,\Gamma(\theta_0-\theta_0')\right]
+O(\varepsilon^2).
\label{map2}
\end{align}
Since the same derivation applies on any interval $[t - \pi,t+\pi]$, and by using $\Gamma(\theta(t-\pi) - \theta'(t-\pi)) = \Gamma (\theta(t) - \theta'(t)) + O(\eps)$, we obtain 
\begin{align}
\theta(t+\pi)-\theta(t-\pi)
=
2\pi\left[1+\varepsilon\,\Gamma\!\left(\theta(t)-\theta'(t)\right)\right]
+O(\varepsilon^2).
\label{map}
\end{align}
Thus, even though the generalized phase does not yield a closed continuous-time equation at leading order, its stroboscopic dynamics closes and depends only on the phase difference. 
We emphasize that the coupling function is the same as in the asymptotic phase map and does not depend on the choice of generalized phase.

\section{Extensions}
\label{extensions}
The above framework admits several extensions; we briefly outline the most relevant ones.

\paragraph{Weak heterogeneity.}
So far, we have considered the dynamics of a generalized phase for a single oscillator.
A natural question is whether the same generalized phase can be used to describe a network of coupled oscillators in a unified manner. 
Consider a system of weakly coupled oscillators with weak heterogeneity, 
\begin{align}
\frac{d\bm X_i}{dt}
=
\bm F(\bm X_i)
+
\varepsilon\,\Delta\bm F_i(\bm X_i)
+
\varepsilon\,\bm \sum_{j=1}^N \bm{P}_i(\boldsymbol{X}_i, \boldsymbol{X}_j), 
\label{osci_i}
\end{align}
where both the heterogeneity and the coupling are of order $O(\eps)$. 
By an analysis parallel to that presented so far, a circle-map description can also be derived for this system. 

Let $\Phi$ and $\Theta$ denote the asymptotic-phase field and generalized-phase field of the reference oscillator \eqref{oscillator}, respectively. 
We define the asymptotic phase of oscillator $i$ as $\phi_i = \Phi(\boldsymbol{X}_i)$ and its generalized phase as $\theta_i = \Theta(\phi_i, a_i)$
The quantities $\psi_i$ and $a_i$ are then introduced in the same manner as in \eqref{expansion}. They follow similar equations as \eqref{dot_psi} and \eqref{dot_a} except that the driving term ~\eqref{G} being replaced by 
\begin{align}
\bm G_i(t)
=
S^{-1}(t+\phi_{i,0})
\left\{
\Delta\bm F_i\!\left(\bm X_0(t+\phi_{i,0})\right)
+
\sum_j \bm P_i( \boldsymbol{X}_0(t+\phi_{i,0} ),\boldsymbol{X}_0(t+\phi_{j,0} )  )
\right\}.
\label{G2}
\end{align}
The corresponding function $H$ in~\eqref{H} is replaced by $H_i$, obtained from $\psi_i$ and $a_i$. Following the same derivation, we obtain the circle map
\begin{align}
    \frac{\theta_i(t+\pi) - \theta_i(t-\pi)}{2\pi} = 1 + \eps \Delta \omega_i + \eps \sum_j \Gamma_i (\theta_i - \theta_j),
\end{align}
where
\begin{align}
\Delta \omega_i
=
\frac{1}{2\pi}
\int_{0}^{2\pi} ds\;
\bm u_0\cdot S^{-1}(s)\,\Delta\bm F_i(\bm X_0(s))
\end{align}
is the effective frequency shift.


\paragraph{Higher-order interactions.}
The framework can also incorporate higher-order (multi-body) interaction terms, e.g., three-body contributions
\begin{align}
    \frac{\theta_i(t+\pi) - \theta_i(t-\pi)}{2\pi} = 1 + \eps \Delta \omega_i + \varepsilon
\sum_{j,k}
\sum_{l_1+l_2+l_3=0}
Q_{i j k l_1 l_2 l_3}\!\left(l_1\theta_i+l_2\theta_j+l_3\theta_k\right),
\end{align}
where $l_1,l_2,l_3$ are integers satisfying $l_1+l_2+l_3=0$.


\section{Coupling inference based on the circle map}
\label{sec:couling-inf}
An important application of the circle map is the inference of interactions between oscillators. 
In many previous studies (e.g., ~\cite{tokuda2019practical, ota2020interaction, novaes2021recovering}), such inference has been performed based on continuous-time phase equations, such as Eq. \eqref{contin-model}. As we have shown, the usage of continuous-time phase models is valid if we can reconstruct the asymptotic phase from trajectory data. 
However, in practice, obtaining the asymptotic phase is often difficult, and it is generally unclear to what extent an empirically obtained phase approximates it. 
In contrast, the generalized phase introduced here is more readily accessible, and we have shown that its dynamics is governed, at leading order, by a circle map with the same coupling function regardless of the particular phase definition. 
This result motivates an inference method based on the circle map \eqref{map_th}, which estimates oscillator coupling from observed trajectories via readily computable generalized phases.
We compare its performance with a conventional method based on the continuous-time model \eqref{contin-model}.

\subsection{Proposed method}
We present a coupling inference method based on the circle map. 
Here, we consider two coupled oscillators, and describe a procedure for inferring the coupling function $\Gamma_1$ associated with oscillator 1. The coupling function $\Gamma_2$ for oscillator $2$ can be inferred in the same manner by exchanging subscripts $1$ and $2$.  

First, we estimate the natural frequency, for example, by counting the number of times the oscillator crossings a certain section, as detailed in Sec.~\ref{application-vdp}. 
Next, we rescale time and denote the rescaled time by $t$, so that the natural period of oscillator $1$ becomes $2\pi$. 
We then use phase values at half-period intervals, $\{ \theta_i(n\pi)\}_{n=0,1,...,N_\pi} $ ($i=1,2$), which may be obtained from the observed phase time series by interpolation if necessary.  

From the circle map \eqref{map_th}, $\Gamma_1$ satisfies
\begin{align}
    \Gamma_1(\theta_1(t) - \theta_2(t)) = \frac{ \theta_1(t+\pi) - \theta_1(t-\pi) - 2\pi }{2 \pi \eps } \label{Gamma-map}
\end{align}
where $O(\eps)$ terms are neglected. 
The coupling function is expanded in a Fourier series as 
\begin{align}
    \Gamma_1(\psi; \boldsymbol{a}, \boldsymbol{b}) = a_0 + \sum_{k=1}^K (a_k \cos k \psi + b_k \sin k \psi), \label{Q-FT}
\end{align}
where $\boldsymbol{a}=(a_0, a_1, ..., a_K), \boldsymbol{b}=(b_1, b_2, ..., b_K)$, and $K$ determines the cutoff frequency.
We estimate the Fourier coefficients $\boldsymbol{a}$ and $  \boldsymbol{b}$ by least squares, i.e., by minimizing
\begin{align}
    \sum_{n=1}^{N_\pi - 1} \left(  \Gamma_1(\theta_1(n\pi) - \theta_2(n\pi); \boldsymbol{a}, \boldsymbol{b}) -   \frac{ \theta_1((n+1)\pi) - \theta_1((n-1)\pi) - 2\pi }{2 \pi \eps }\right)^2.
\end{align}

\subsection{Conventional method}
For comparison, we consider a conventional method based on the continuous-time phase model given by \eqref{contin-model}.
Again, we describe a procedure for inferring the coupling function $\Gamma_1$ associated with oscillator 1. 
We assume that the time is rescaled so that the natural period of oscillator $1$ becomes $2\pi$, and phase data sampled at a small time step $\Delta t$, $\{ \theta_i(n\Delta t)\}_{n=0,1,...,N_{\Delta t}} $ ($i=1,2$) are available. 

By discretizing \eqref{contin-model} using the forward Euler scheme and rearranging it, we have
\begin{align}
    \Gamma_1(\theta_1(t) - \theta_2(t)) = \frac{ \theta_1(t+\Delta t) - \theta_1(t) - \Delta t }{\eps \Delta t}. \label{Gamma-contin}
\end{align}
We estimate the Fourier coefficients $\boldsymbol{a}$ and $  \boldsymbol{b}$ in \eqref{Q-FT} by least squares, i.e., by minimizing
\begin{align}
    \sum_{n=0}^{N_{\Delta t} - 1} \left(   \Gamma_1(\theta_1(n\Delta t) - \theta_2(n\Delta t); \boldsymbol{a}, \boldsymbol{b}) -   \frac{ \theta_1((n+1)\Delta t) - \theta_1(n \Delta t) - \Delta t }{ \eps \Delta t}\right)^2.
\end{align}

\subsection{Application to van der Pol oscillators}
\label{application-vdp}
We employ a pair of coupled van der Pol oscillators
\begin{subequations}
    \label{vdp}
    \begin{align}
        \tau_i^{-1}\frac{dx_i}{d\tilde{t}} &= - y_i, \\
        \tau_i^{-1}\frac{dy_i}{d\tilde{t}} &= - \mu (1-x_i^2) y_i + x_i - c_i ( y_j - y_i ),
    \end{align}
\end{subequations}
where $i, j \in \{1, 2\} \ (i \neq j)$ are the oscillator indices, $\mu$ controls the strength of the nonlinear damping, $c_i=O(\eps)$ is the small coupling strength, and $\tau_i$ is the time scale parameter introduced to tune the oscillation period.
Increasing \(\mu\) enhances the amplitude stability of the limit cycle, whereas it also deforms the limit-cycle trajectory away from a nearly circular shape.
We generate simulated data $(x_i(\tilde{t}), y_i(\tilde{t}))$ for $\tilde{t} \in [0, \tilde{T}_{total}]$ by numerically integrating \eqref{vdp}. We choose the initial condition $(x_i(0), y_i(0)) = (2, 0)$ for $i=1,2$, which is close to the limit cycle. 
This choice allows us to avoid the initial relaxation process toward the limit cycle. 
Throughout this section, the parameters are chosen so that the two oscillators are not synchronous.

Here, we infer the coupling function $\Gamma_1$ associated with oscillator $1$.  
First, we rescale time so that the natural period of oscillator $1$ becomes $2\pi$. 
Note that it is fundamentally impossible to estimate the natural period and the constant component in the coupling separately from a set of time series $(x_i(t), y_i(t)) \ (i=1,2)$. Here, we assume that the constant component of the coupling function vanishes ($a_0 = 0$), and any constant component is absorbed into the natrual frequency.
Therefore, we regard the typical period $\tilde{T} = (t^{(M)} - t^{(1)}) / (M-1)$ as the natural period, where $M$ is the number of times that the oscillator $1$ crosses the section $x_1=0$ from $x_1<0$ to $x_1>0$, and $t^{(1)}$ and $t^{(M)}$ denote the times of the first and last crossings, respectively. 
We then rescale time by 
\begin{align}
    t = \frac{2\pi}{\tilde{T}} \tilde{t},
\end{align}
so that the resulting period becomes $2\pi$, and we obtain $(x_i(t), y_i(t))$ for $t\in [0, T_{total}]$ with $T_{total} = \frac{2\pi}{\tilde{T}} \tilde{T}_{total}$.
Next, we obtain the data used for inference by linearly interpolating the simulated trajectories.
For the continuous-model-based method, the data are sampled at small time intervals $\Delta t = 0.01$, yielding $\{(x_i(n \Delta t), y_i(n \Delta t) )\}_{n=0, 1, ..., N_{\Delta t} } \ (i=1,2)$, where $N_{\Delta t}$ is the largest integer less than $T_{total} / \Delta t$.
For the circle-map-based method, the data are sampled at intervals of $\pi$, yielding $\{(x_i(n \pi), y_i(n \pi) )\}_{n=0, 1, ..., N_{\pi} } \ (i=1,2)$, where $N_{\pi}$ is the largest integer less than $T_{total} / \pi$. 
In the following, the final time in the simulation data is $T_{total} = 500$, which corresponds to approximately $80$ cycles. 

\begin{figure}
    \centering
    \includegraphics[width=\linewidth]{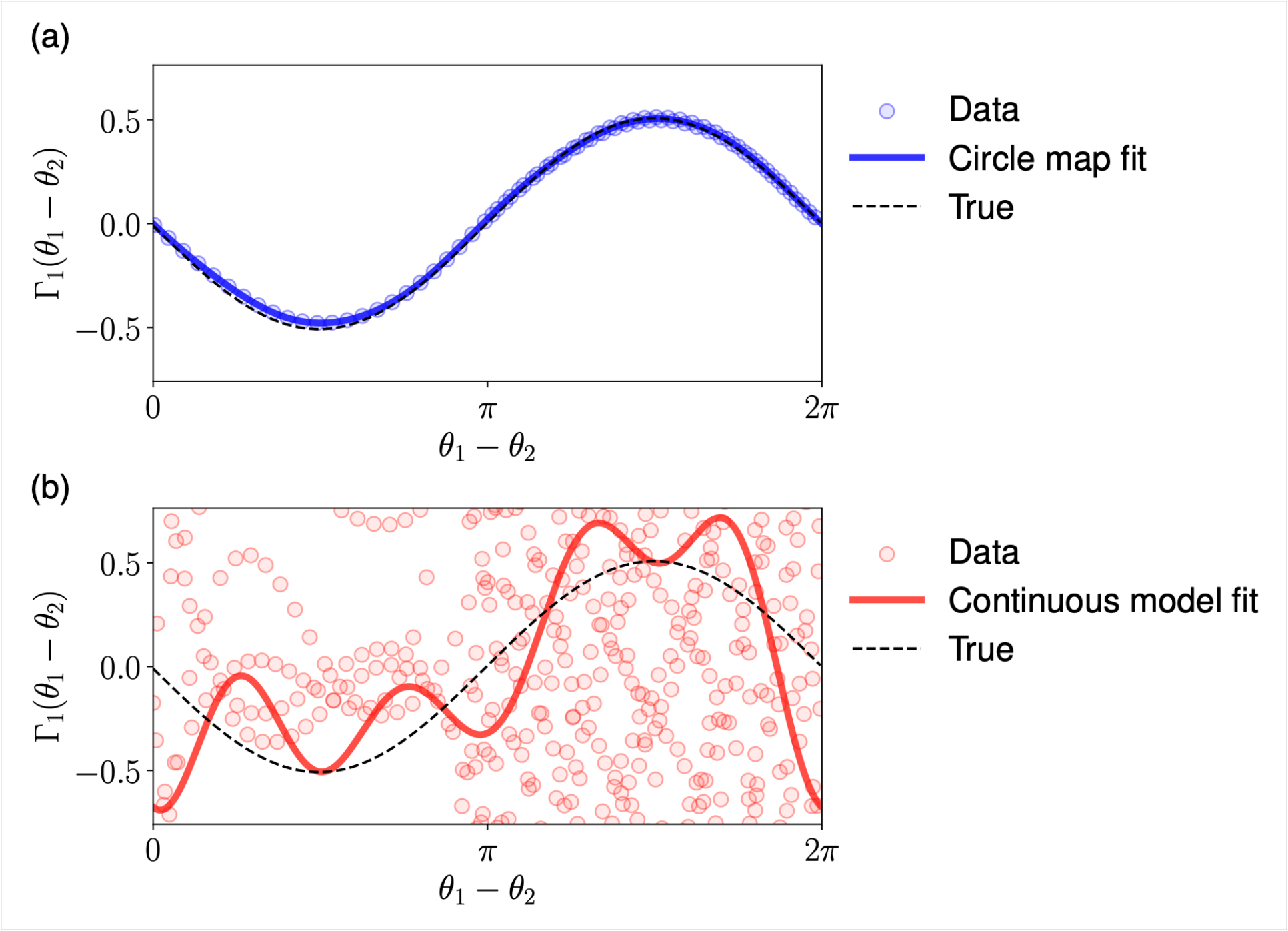}
    \caption{An example of coupling inference from the van der Pol oscillators. 
    (a) Inference based on the circle map. The dots represent the scatter plot of $\left( \theta_1( n \pi ) - \theta_2(n \pi),  \frac{ \theta_1((n+1)\pi) - \theta_1((n-1)\pi) - 2\pi }{2 \pi \eps } \right) \quad (n=1,3,5,...)$. 
    The solid blue line is the coupling function inferred based on the circle map. The dashed line represents the true coupling function of the asymptotic phase equation.
    (b) Inference based on the continuous model. The circles represent the scatter plot of $\left( \theta_1(n \Delta t) - \theta_2(n \Delta t),  \frac{ \theta_1((n+1) \Delta t) - \theta_1(n \Delta t) - \Delta t }{\eps \Delta t} \right) \quad (n=0,100,200,...)$, where $\Delta t= 0.01$. To avoid overcrowding, only every 100th data point is plotted.  
    The solid red line is the coupling function inferred based on the continuous model.
    Parameters: $\mu=0.1, c_1=0.005, c_2=0, \tau_1 = 1, \tau_2 = 1.02$, and $T_{total} = 500$. 
    We chose $K=5$ for the inference. }
    \label{fig3}
\end{figure}

\paragraph{Example of inference.} 
Figure \ref{fig3} shows an example of coupling inference using the polar angle 
\begin{align}
    \theta_j(t)=\arg\left(x_j(t)+\mathrm{i}x_j(t-d)\right),  \quad (j=1,2)
\end{align}
where $\mathrm{i}$ is the imaginary unit. 
The blue circles in Fig.~\ref{fig3}a show the scatter plot of 
\begin{align}
    \left( \theta_1( n \pi ) - \theta_2(n \pi),  \frac{ \theta_1((n+1)\pi) - \theta_1((n-1)\pi) - 2\pi }{2 \pi \eps } \right) \quad (n=1,3,5,...)
\end{align}
corresponding to the right-hand side of \eqref{Gamma-map} associated with the circle map, as a function of the phase difference $\theta_1(t)-\theta_2(t)$. 
The black dashed curve represents the true coupling function $\Gamma_1$ of the asymptotic phase equation (See Appendix \ref{appendix-adjoint} for the computation). 
The blue circles lie almost exactly on the black dashed curve, indicating excellent agreement between the coupling term in the circle map and the true coupling function.  
As a result, the estimated coupling by the proposed method (blue solid line) reproduces the true coupling with high accuracy. 

The red circles in Fig.~\ref{fig3}(b) show the scatter plot of 
\begin{align}
    \left( \theta_1(n \Delta t) - \theta_2(n \Delta t),  \frac{ \theta_1((n+1) \Delta t) - \theta_1(n \Delta t) - \Delta t }{\eps \Delta t} \right) \quad (n=0,1,2,...)
\end{align}
corresponding to the right-hand side of the continuous model \eqref{contin-model} as a function of the phase difference. 
In contrast to the blue circles, the red dots are notably displaced from the black curve, i.e., the true coupling function. 
As a result, the estimated coupling function by the conventional method (solid red line) deviates significantly from the true function. 
We also verified that the proposed method is robust to the choice of $K$, the cutoff harmonics of the Fourier expansion \eqref{Q-FT}, whereas the continuous-model-based inference is highly sensitive to this parameter (see Appendix \ref{appendix-K}).

\begin{figure}
    \centering
    \includegraphics[width=\linewidth]{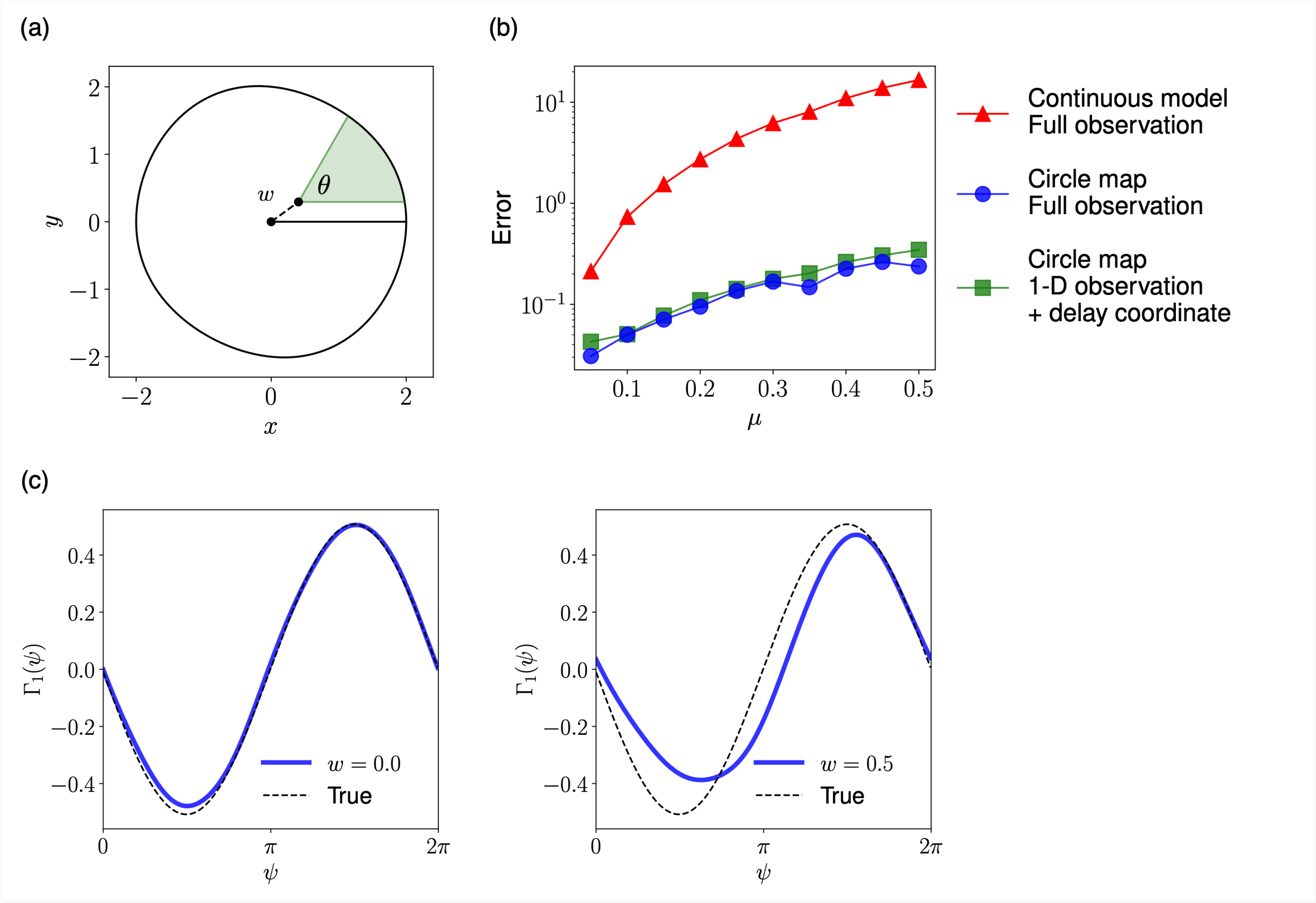}
    \caption{The circle-map-based inference is robust against various choices of the phase. 
    (a) The polar angle with an origin shift of amplitude $v$. 
    (b) The estimation error \eqref{error} as functions of the shift amplitude $w$ from the van der Pol oscillators. The dots and error bars are the mean and the standard deviations over 100 trials with different $\xi$, respectively.
    (c) The examples of inferred coupling functions for $w=0.0$ and $w = 0.5$.
    Parameters: $\mu=0.1, \eps=0.005, c_1=1, c_2=0, \tau_1 = 1, \tau_2 = 1.02$, and $T_{total} = 500$. We chose $K=5$ for the inference. 
    }
    \label{fig4}
\end{figure}

\paragraph{Robustness to the choice of generalized phase.} 
Next, we show that the proposed method is robust against various choices of the generalized phase. 
Here, we construct the generalized phase as the polar angle computed with a shifted origin (Fig.~\ref{fig4}a): 
\begin{align}
    \theta_i(t) = \arctan \left[ \frac{ y_i(t) - y_0}{ x_i(t) - x_0 } \right],
\end{align}
where $(x_0, y_0)$ denotes the shifted origin, chosen randomly as 
\begin{align}
    x_0 + i y_0 = w \exp(i \xi)
\end{align}
with $\xi$ drawn from the uniform distribution on $[0, 2\pi)$. 
As the shift amplitude $w$ increases, the phase velocity on the limit cycle of the unperturbed oscillator deviates from constant, thus the assumption \eqref{assumption2} becomes less accurate.
Figure \ref{fig4}b shows the inference error 
\begin{align}
    {\rm Error} =  \sqrt {\frac{\int_0^{2\pi} d\psi \left( \hat{\Gamma}_1(\psi) -\Gamma_1(\psi) \right)^2 }{\int_0^{2\pi} d\psi \left( \Gamma_1(\psi) \right)^2 }} \label{error}
\end{align}
as a function of the shift amplitude $w$.
The circle-map-based inference is more robust to the shift of origin compared to the continuous-model-based inference. 
Notably, even for $w=0.5$, corresponding to a shift whose magnitude is approximately 25\% of the oscillation amplitude, the inference error is only around 0.1, and the inferred coupling function still agrees reasonably well with the true one (Fig.~\ref{fig4}c)

\paragraph{Robustness to distortion of the limit cycle and weak transverse stability.} 
Furthermore, we examine the robustness of the proposed method by varying the van der Pol parameter $\mu$, which controls two competing properties relevant to our assumptions. 
For small $\mu$, the limit-cycle trajectory is close to circular and the polar angle advances nearly uniformly, whereas amplitude relaxation toward the limit cycle is relatively weak. 
For large $\mu$, amplitude relaxation becomes stronger, but the limit cycle is increasingly distorted from a circle and the polar angle rotates less uniformly along the unperturbed cycle. 
Thus, varying $\mu$ allows us to test the proposed method against both weak amplitude stability and deviations from near-uniform phase rotation.

\begin{figure}
    \centering
    \includegraphics[width=\linewidth]{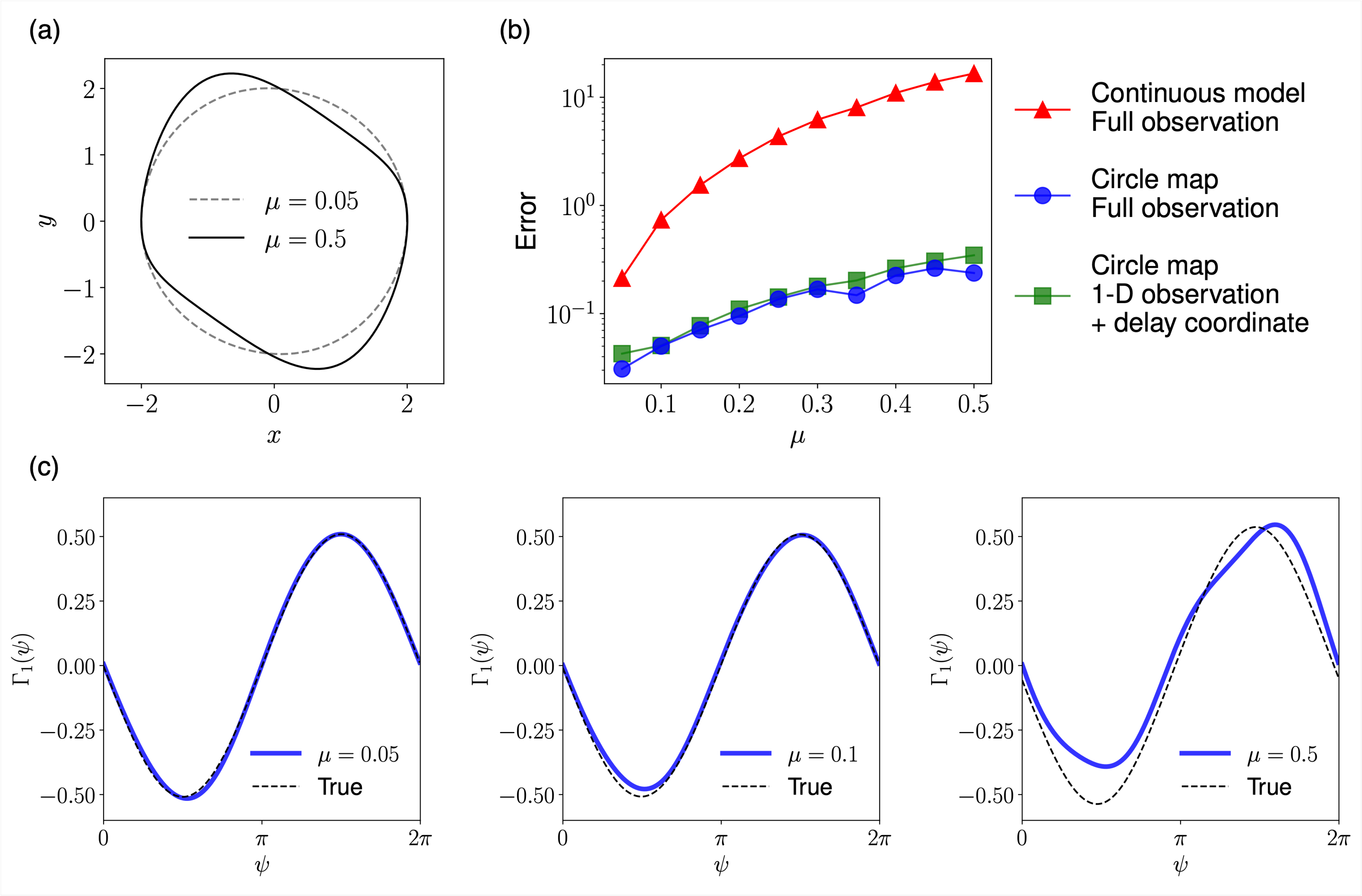}
    \caption{The circle-map-based inference is robust against noncircular limit cycles.
    (a) The limit cycles with $\mu=0.05$ and $\mu=0.5$. 
    (b) The estimation error \eqref{error} as functions of $\mu$. 
    (c) The examples of inferred coupling functions for $\mu=0.05, 0.1,$ and $0.5$.
    Parameters: $\eps=0.005, c_1=1, c_2=0, \tau_1 = 1, \tau_2 = 1.02$, and $T_{total} = 500$. We chose $K=5$ for the inference. 
    }
    \label{fig5}
\end{figure}

As shown in Fig.~\ref{fig5}a, the limit cycle is nearly circular for small $\mu$, such as $\mu=0.05$, and becomes increasingly noncircular as $\mu$ increases. 
Despite the weak amplitude relaxation at $\mu=0.05$, Fig.~\ref{fig5}b and Fig.~\ref{fig5}c(left) show that the inference remains accurate in this regime. 
This suggests that the circle-map-based inference can remain accurate even when amplitude relaxation is not very strong, provided that the generalized phase rotates nearly uniformly. 
For larger $\mu$, the inference error remains small (Fig.~\ref{fig5}b) and the inferred coupling function still agrees reasonably well with the true one (Fig.~\ref{fig5}c) even though the limit cycle is substantially distorted and the near-uniform-rotation assumption becomes less accurate. 
These results indicate that the proposed method is robust over a broad range of $\mu$, including regimes in which either amplitude relaxation is weak or the near-uniform-rotation assumption is less accurately satisfied.

\paragraph{Inference from delay-coordinate-based phase.} 
Finally, we examine the case in which only a single observable is available.
Assuming that we observe only $x_1(t)$ and $x_2(t)$, we reconstruct the phase-like variable using delayed coordinates:
\begin{align}
    \theta_i(t) = \arctan \left[ \frac{x_i(t - d)}{x_i(t)} \right], \quad (i=1,2)
\end{align}
where $d=\pi/2$ is chosen as a quater of the typical period $2\pi$.
Striclty speaking, this $\theta_i$ does not correspond to the generalized phase considered in our theory, because it is not a function of the instantaneous state vector $(x_i,y_i)$. Nevertheless, it can serve as a practical phase-like approximation constructed solely from a one-dimensional observation. 
As in the previous case, we examine the accuracy of the inference for various values of $\mu$. 
The green line in Fig.~\ref{fig5}b shows that the inference error is comparable to that obtained using the polar angle. 
This result suggests that the circle-map-based inference remains applicable even when 
the available phase variable is an approximate phase reconstructed from a one-dimensional oscillatory signal.

\section{Conclusion and Discussion}
\label{conclusion}
In this paper, we analyzed the dynamics of generalized phases, which are defined as smooth functions on the state space and are generally much easier to construct from observed data than asymptotic phases. 
We showed that, in a continuous-time description, amplitude deviations generally contaminate the phase dynamics, preventing the resulting equation from being closed in terms of the phase alone. 
To overcome this difficulty, we considered the stroboscopic dynamics of generalized phases and derived a simple and unified description in the form of a circle map under mild conditions. 

Importantly, we showed that our circle map is invariant to 
the choice of generalized phase as far as the assumptions \eqref{assumption1} and \eqref{assumption2} are met; the coupling term is the same as that for the asymptotic phase, which is a one-variable function of the phase difference. 
We numerically verified this fact using the Stuart-Landau oscillators (see Fig.~\ref{fig2}).


This property makes the circle map particularly useful for coupling inference from oscillatory data: 
by fitting phase data to the circle map, one can infer the coupling function independent of the choice of phase. 
We evaluated the performance of the circle-map-based inference using simulated data from coupled van der Pol oscillators. 
The proposed method achieved high accuracy for a wide range of phase choices, and also even for cases where the limit-cycle trajectory deviated substantially from a circular orbit. 
We also demonstrated numerically that accurate inference is possible even when only a one-dimensional signal is available, by reconstructing the state space using delay coordinates. Although phases obtained using delay coordinates are not generalized phases employed in our theory, the results suggest that the circle-map-based framework may remain effective beyond the class of phase variables covered by the present analysis.

In this study, we considered only two-dimensional oscillator systems in our numerical experiments. However, the theoretical framework is not restricted to two dimensions, and the same circle-map description is expected to hold for higher-dimensional systems, as long as the system's dynamics settle in the vicinity of the limit cycle with sufficiently strong orbital stability.
Applying the proposed method to such systems therefore constitutes an important direction for future research. Even for high-dimensional oscillator systems, it is expected to suffice to observe only two state variables to construct a geometric-angle phase, which has been a successful heuristic in identifying phase dynamics in coupled oscillators\cite{rosenblum1996phase, pikovsky2003synchronization}
When only a one-dimensional signal can be observed from each oscillator, the phase reconstruction via delay coordinate \cite{packard1980geometry, takens1981detecting, rosenblum1996phase, pikovsky2003synchronization} or via the Hilbert transform\cite{delprat1992, cohen1999ambiguity, chavez2006} or recent advanced methods \cite{gengel2019phase, gengel2022phase, matsuki2023extended} may provide an alternative approach.

Another promising direction is to investigate other types of perturbations, as briefly discussed in Sec.~\ref{extensions}. In particular, extending the framework to higher-order interactions may be of considerable interest. Whereas classical phase-reduction approaches typically consider only pairwise interactions, recent studies have highlighted the importance of higher-order interactions involving three or more oscillators \cite{battiston2020networks, bick2023higher, boccaletti2023structure}. Corresponding methods for detecting and inferring such interactions have also been developed \cite{kralemann2014reconstructing, stankovski2015coupling, casadiego2017model, pikovsky2022non, rosenblum2023inferring, neuhauser2024learning, malizia2024reconstructing, delabays2025hypergraph,su2025distinguishing}. 
Because our framework focuses on stroboscopic phase dynamics rather than continuous-time dynamics, it may provide a useful basis for higher-order-interaction inference from data-driven phase representations.

Finally, although this work focused on coupling inference for asynchronous systems, inference in synchronous systems is also an important problem\cite{mori2022noninvasive, matsuki2025network}. Such inference is known to be challenging because nearly phase-locked oscillators provide only limited information about the underlying coupling\cite{rosenblum2001detecting, tokuda2007inferring, tokuda2019practical}. One possible strategy is to exploit fluctuations around the phase-locked state. 
In \cite{matsuki2025network}, it has been demonstrated that a circle-map-based method performs well for noisy synchronous oscillators. 
In that work, however, the theory based on the circle map was established only for asymptotic phases, leaving it unclear whether phases reconstructed from data obey the same description. The present study provides a theoretical justification for applying circle-map-based inference to reconstructed generalized phases.
Applying this framework to real-world data (e.g., neural systems~\cite{Friston1994, messe2015closer, barack2022call, kobayashi2025inference}, the suprachiasmatic nucleus (SCN)~\cite{myung2015gaba} and spinal cord \cite{kobayashi2016}) is an important future direction, as experimental systems inevitably involve observational and system noises, limited data availability, and possible deviations from the assumptions of phase models. Such studies will be essential for assessing the practical utility of coupling inference in oscillator networks.

\






\appendix
\section{Appendix: Floquet theory}
\label{appendix-floquet}
\addcontentsline{toc}{section}{Appendix: Floquet theory}

We briefly summarize Floquet theory for the linearized dynamics around a stable limit cycle of the unperturbed system~\eqref{oscillator}.
Let $\bm X_0(t)$ be the $T$-periodic limit-cycle solution ($T=2\pi$ in this paper).
Linearizing~\eqref{oscillator} about $\bm X_0(t)$ yields the variational equation
\begin{align}
\frac{d\bm z}{dt} = L(t)\bm z,
\label{variational}
\end{align}
where $L(t)$ is the Jacobian matrix along the cycle,
\begin{align}
L(t)=\left.\frac{\partial \bm F}{\partial \bm X}\right|_{\bm X=\bm X_0(t)}.
\end{align}
Since $\bm X_0(t)$ is $T$-periodic, $L(t)$ is also $T$-periodic: $L(t+T)=L(t)$.

Floquet theory states that the fundamental matrix solution of~\eqref{variational} can be written as
\begin{align}
\bm z(t)=S(t)e^{\Lambda t}\bm c,
\label{floquet_form}
\end{align}
where $S(t)$ is a $T$-periodic, invertible matrix ($S(t+T)=S(t)$), $\Lambda$ is a constant matrix, and $\bm c$ is a constant vector determined by the initial condition.
Differentiating~\eqref{floquet_form} and using~\eqref{variational}, we obtain the identity
\begin{align}
\dot S(t)+S(t)\Lambda = L(t)S(t).
\label{floquet_id}
\end{align}

We assume that $\Lambda$ is diagonalizable.
Let $\{\lambda_i\}_{i=0}^{N-1}$ be its eigenvalues, with corresponding right and left eigenvectors $\bm v_i$ and $\bm u_i$:
\begin{align}
\Lambda \bm v_i &= \lambda_i \bm v_i,\\
\bm u_i \Lambda &= \lambda_i \bm u_i,
\end{align}
where $N$ is the dimension of the system (here $N=2$ in Sec.~3).

Because the limit cycle is autonomous, there is a neutral direction corresponding to time translation.
Accordingly, one Floquet exponent is zero; we denote it by $\lambda_0=0$.
The associated right eigenvector can be chosen as the tangent vector to the orbit at $t=0$:
\begin{align}
\bm v_0 = \frac{d{\bm X}_0(0)}{dt}=\bm F(\bm X_0(0)).
\end{align}
All remaining Floquet exponents satisfy $\mathrm{Re}(\lambda_i)<0$ for a stable limit cycle.

By an appropriate normalization of left and right eigenvectors, we may assume a biorthonormality relation
\begin{align}
\bm u_i \cdot \bm v_j = \delta_{ij}.
\label{biorth}
\end{align}
This biorthonormal system underlies the mode decomposition used in Sec.~3: deviations from the limit cycle are expanded in the neutral (phase) mode and the stable transverse (amplitude) mode, and the corresponding coefficients are obtained by projection with the left eigenvectors.

In the planar case ($N=2$), there is exactly one stable transverse mode with exponent $\lambda_1=-\lambda$ (with $\lambda>0$), which we use as the amplitude direction in Sec.~3.

\section{Numerical computation of the true coupling function}
\label{appendix-adjoint}
To obtain the true phase coupling function $\Gamma_1$ of the van der Pol oscillator, we first numerically compute the limit cycle solution $(x^{lc}(t), y^{lc} (t) )$ for $t \in [0, 2\pi)$ (the time is rescaled so that the period becomes $2\pi$) of the unperturbed oscillator. 
We also compute the phase sensitivity function $\boldsymbol{Z}(s), \ s \in [0, 2\pi)$ with the adjoint method \cite{ermentrout1996type, hoppensteadt1997, ermentrout2010mathematical}. 
Then, we compute 
\begin{align}
    \Gamma_1(\psi) = \int_0^{2\pi} ds \boldsymbol{Z}(s) \cdot \boldsymbol{p}_1(s, s - \psi),
\end{align}
where 
\begin{align}
    \boldsymbol{p}_1(s_1, s_2) = \left(\begin{array}{c}
        0  \\
        -\tau_1 c_1 y^{lc}_2(s_2) 
    \end{array} 
    \right) \quad (s_1, s_2 \in [0, 2\pi))
\end{align}
is the coupling to the other oscillator in the original dynamical system. 
Note that we do not include the self-coupling term $\tau_1 c_1 y_1^{lc}$ in the above, because it yields a constant term $a_0$ in $\Gamma_1$, which we assume to be zero. Instead, the contribution from the self-coupling term appears in the typical frequency. 
$\Gamma_2$ can also be computed in the same manner.

\section{Robustness of the coupling inference with respect to the cutoff frequency}
\label{appendix-K}
\begin{figure}
    \centering
    \includegraphics[width=\linewidth]{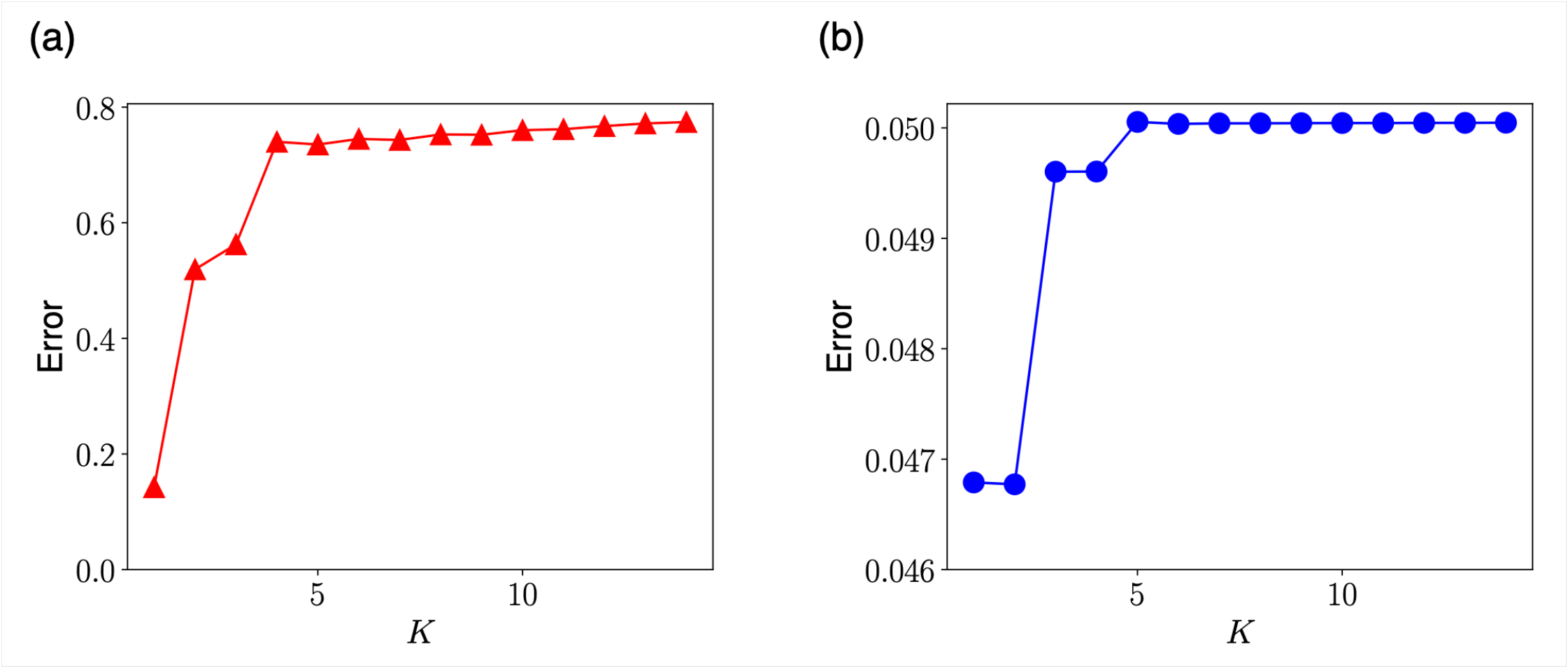}
    \caption{The circle-map-based inference is robust with respect to the cutoff frequency.  
    The error \eqref{error} of the coupling inference (a) by the continuous-model-based method and (b) by the circle-map-based method are plotted as a function of $K$.}
    \label{fig-appendix}
\end{figure}

Figure~\ref{fig-appendix} shows the estimation error \eqref{error}
of the coupling inference for the coupled van der Pol oscillators \eqref{vdp} as a function of $K$ in \eqref{Q-FT}, which specifies the cutoff frequency in the Fourier expansion of the coupling function. 
For the continuous-model-based inference, increasing $K$ substantially increases the estimation error, indicating overfitting due to spurious higher-harmonic components (Fig. \ref{fig-appendix}a). 
In contrast, the estimation error of the circle-map-based inference remains very small as $K$ increases (Fig. \ref{fig-appendix}b). These results demonstrate that the proposed method is robust to the choice of $K$.

\end{document}